\documentclass[aps,prl,twocolumn,groupedaddress]{revtex4}
\usepackage{graphicx}
\usepackage{amsmath}
\usepackage{natbib}

\newcommand{\topen}{{\tau_{\rm a}}}   
\newcommand{\tclose}{{\tau_{\rm i}}}  
\newcommand{\kopen}{{k_{\rm u}}}        
\newcommand{\kclose}{{k_{\rm b}}}       
\newcommand{\Keq}{{K}}                  
\newcommand{\Uchain}{{U_{\rm s}}}   
\newcommand{\Ubend}{U_{\rm b}}  
\newcommand{\UDH}{U_{\rm DH}}      
\newcommand{\Uc}{U_{\rm c}}        
\newcommand{\rvec}{{\bf r}}        
\newcommand{\cvec}{{\bf c}}        
\newcommand{\chainstiff}{\varepsilon_{\rm s}}  
\newcommand{\bendstiff}{\varepsilon_{\rm b}}   
\newcommand{\etavec}{\mbox{\boldmath$\eta$}}   
\newcommand{\lp}{\ell_{\rm p}}     
\newcommand{\lc}{\ell_{\rm c}}     
\newcommand{\mwt}{\tau_{\rm w}}     
\newcommand{\uev}{\lambda_{-}}     
\newcommand{\kruem}{\Gamma}     
\newcommand{\Lcross}{\ell_{\rm c}}     

\newcommand{\FIG}[1]{Fig.~\ref{fig:#1}}

\begin{document}

\title{Kinetic accessibility of buried DNA sites in nucleosomes}

\author{Wolfram M\"obius}
\author{Richard A. Neher} 
\author{Ulrich Gerland}

\affiliation{
Arnold Sommerfeld Center for Theoretical Physics (ASC) and Center for Nanoscience (CeNS), LMU M\"unchen, Theresienstr. 37, 80333 M\"unchen, Germany}

\date{\today}

\begin{abstract} 
Using a theoretical model for spontaneous partial DNA unwrapping from histones,
we study the transient exposure of protein-binding DNA sites within
nucleosomes. We focus on the functional dependence of the rates for site
exposure and reburial on the site position, which is measurable experimentally
and pertinent to gene regulation. We find the dependence to be roughly
described by a random walker model. Close inspection reveals a surprising
physical effect of flexibility-assisted barrier crossing, which we characterize
within a toy model, the `Semiflexible Brownian Rotor'. 
\end{abstract}

\pacs{???}
\maketitle

\begin{figure}[b] 
\includegraphics[width=8.2cm]{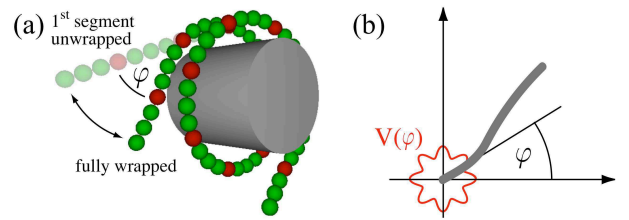}
\caption{\label{fig1}
(a) Illustration of our nucleosome model. The DNA-histone interaction is localized at contact points 
attracting the red (dark) beads. The DNA is shown in the ground state as well as a conformation where the 
first contact is open. 
(b) Illustration of the Semiflexible Brownian Rotor (SBR) model. In this toy model, the tradeoff between bending energy and DNA-histone interaction in the nucleosome is mimicked by an angular potential $V(\varphi)$, exerting a torque on the attachment angle $\varphi$ of a semiflexible polymer at the origin.} 
\end{figure}

Although the DNA in eukaryotic cells is packaged into chromatin, its genetic information must be 
accessible to proteins for read out and processing \cite{Luger:05}. The structural organization of 
chromatin is fairly well known: the fundamental unit is a nucleosome core particle (NCP) consisting of 
about 150 base pairs (bp) of DNA wrapped in $1.7$ turns around a cylindrical histone octamer \cite
{Luger:97}, and NCPs are regularly spaced along the DNA, which is further compactified into higher 
order structures. In contrast, the {\em conformational dynamics} of chromatin is poorly understood. 
Recent experiments studied these dynamics on the level of individual NCPs using single-molecule force 
\cite{Brower-Toland:02} and fluorescence \cite{Li:05,Tomschik:05} techniques. The latter directly 
observed spontaneous conformational transitions where part of the DNA unwraps reversibly, allowing 
proteins to access DNA sites that are normally buried. This mode of access, driven by thermal 
fluctuations, is particularly important for passive DNA-binding proteins, e.g., transcription factors. 
Here, we study spontaneous DNA unwrapping within a theoretical model, see Fig.~\ref{fig1}(a). 

Consider a buried DNA site that is accessible only when a DNA segment of length $L$ is unwrapped. How 
long is the typical dwell time $\topen$ in the accessible state, i.e., the window of opportunity for 
protein binding? And what is the typical time $\tclose$ for which it remains inaccessible? Li {et al.} 
\cite{Li:05} measured $\topen\!=\!10\!-\!50$~ms and $\tclose\!\approx\!250$~ms for $L\!\sim\!30$~bp, 
while Tomschik {et al.} \cite{Tomschik:05} found $\topen\!=\!100\!-\!200$~ms and $\tclose\!=\!2\!-\!5
$~s for $L\!\sim\!60$~bp. Taken together, these results indicate a significant dependence on $L$ in 
both time scales, which cannot be reconciled with an early theoretical study \cite{Marky:95} suggesting 
an all-or-none unwrapping mechanism where the nucleosome fluctuates between two conformations only. 
Instead, these results, as well as previous biochemical experiments
\cite{Polach:95}, imply a multistep opening mechanism. 

In this Letter, we propose and characterize a theoretical model for this
multistep mechanism, similar in spirit to previous work on histone-DNA
interactions which focused mainly on static properties or the calculation of
free energy barriers \cite{Marky:95,nuc_theory,Schiessel:03}. Within our model,
we clarify the physics that determines the $L$ dependence of the time scales
$\topen$ and $\tclose$. We find that the dependence of $\tclose$ can be
interpreted with a simple random walker model, which may serve as a fitting
model for future experiments that probe the time scales at different $L$
values. In contrast, the $L$ dependence of $\topen$ reflects the intricate
coupling between the DNA polymer dynamics and the dynamics of breaking and
reforming DNA-histone contacts. To analyze the effect of this coupling, we
introduce a toy model, the Semiflexible Brownian Rotor (SBR), see
Fig.~\ref{fig1}(b). We identify a generic physical effect of
flexibility-assisted barrier crossing, which may arise also in other contexts.
It is marked by a characteristic plateau of the time scale at intermediate $L$.
Biologically, the $L$ dependence is relevant, because it creates a positioning
effect for transcription factor binding sites relative to nucleosomes
\cite{Segal:06}. We expect that the integration of single NCPs into nucleosome
arrays will alter the {\em absolute} time scales but not the basic physics of
the DNA (un)wrapping process.

{\it Nucleosome model.---} 
The NCP crystal structure \cite{Luger:97} shows that both the electrostatic and hydrogen bond 
interactions between the DNA and the histone complex are mainly localized to 14 contact points, about 
evenly spaced by 10.2~bp along a superhelical contour with radius 4.2~nm and helical pitch 2.4~nm.
Because we are interested only in the dynamics at a fixed (physiological) salt 
concentration, we combine the interactions at each of these points into a simple Morse potential \cite
{note4}. The DNA-histone interaction energy is then 
\begin{equation}
\label{contact_potential}
  \Uc = \gamma\, k_BT \sum_{n}\big(1-e^{-|\rvec_{i(n)}-\cvec_n|/\rho}\big)^2 \;,
\end{equation}
where $\cvec_n$ is the $n$th contact point on the superhelical contour,
$\gamma$ is the depth, and $\rho$ the width of the contact potential. A
discrete bead-spring model with beads at positions $\rvec_i$ models the DNA,
and $i(n)$ is the bead bound to contact $n$ in the fully wrapped state. The
beads are connected by a harmonic potential $\Uchain\!=\!\chainstiff\sum_i
\left(\left|\rvec_{i+1}\!-\!\rvec_i\right|\!-\!a\right)^2\!/2$ with a typical
bead separation $a$ and a stiffness $\chainstiff$ set to $800\,k_{B}T/{\rm
nm}^2$. Below, we use 3 beads between contacts and at each end (about
$2.5\,$bp/bead), unless stated otherwise. Increasing the discretization or
$\chainstiff$ raises the computational effort without affecting our results
qualitatively. We account for the bending rigidity of DNA by an energy
$\Ubend=\bendstiff\sum_{i} \left(1-\cos{\theta_i}\right)$ with bending angle
$\theta_i$ at bead $i$ and a bending stiffness $\bendstiff$ adjusted such that
the apparent persistence length matches the known $\lp\approx 50$~nm for DNA at
physiological salt conditions. Furthermore, we incorporate the screened
electrostatic self-repulsion of DNA through a Debye-H\"uckel potential $\UDH =
k_BT\,l_B(\tau a)^2\,\sum_{i<j}
e^{-\kappa|\rvec_i-\rvec_j|}/|\rvec_i\!-\!\rvec_j|$ with the Bjerrum length
$l_{B}\approx 0.7$~nm, a charge density $\tau=2$ charges/bp, and a screening
length $\kappa^{-1}\approx 1$~nm. We use a contact radius $\rho=0.5$~nm in
between the range of hydrogen bonds and electrostatic interactions and adjust
the depth $\gamma$ of the Morse potential to match the binding free energy
\cite{note2} of $\approx\! 1.5\,k_{B}T$ per contact estimated from biochemical
experiments \cite{Polach:95,Schiessel:03}. Taken together, the total energy is
$U = \Uchain + \Ubend + \UDH + \Uc$. To study the dynamics of our model, we
perform Brownian dynamics simulations with the overdamped Langevin Eqs. 
\begin{equation}
  \label{Langevin}
  \dot{\rvec}_{i}(t) = -\mu_{b}\nabla_{\rvec_{i}}U(\{\rvec_{j}\})+\etavec_{i}(t) \;,
\end{equation}
where $\mu_{b}$ is the bead mobility, and the absolute time scale is set by $a^2/\mu_{b}k_{B}T$. The random forces $\etavec_{i}$ satisfy $\langle\etavec_{i}(t)\cdot\etavec_{j}(t')\rangle=6\,\mu_{b} k_{B}T\,\delta_{i,j}\delta(t-t')$.

\begin{figure}[t]
\includegraphics[width=8.5cm]{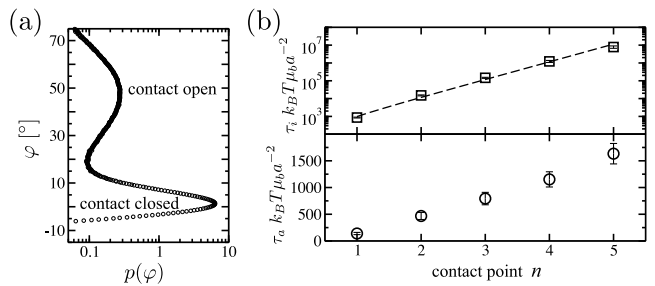}
\caption{\label{fig2} 
(a) Equilibrium distribution of the DNA angle $\varphi$ defined in Fig.~\ref{fig1}(a). The two peaks at $\varphi=0$ and $\varphi\approx 45^{\circ}$ correspond to the fully wrapped state and the state with contact 1 open, respectively. 
(b) Kinetics of DNA site exposure within our nucleosome model. The dwell time in the inaccessible state (squares) increases roughly exponentially with the number of contacts that must open to render a DNA site accessible. The dashed line is a fit to Eq.~(\ref{MFPT}). The circles show the average time the $n$th contact point remains open.} 
\end{figure}

{\it Unwrapping dynamics.---} 
A suitable reaction coordinate for the opening of a single contact is the
attachment angle $\varphi$, see Fig.~\ref{fig1}(a), which changes by
$\Delta\varphi\approx45^{\circ}$ in this process. The equilibrium distribution
$p(\varphi)$ for the first contact is shown in Fig.~\ref{fig2}(a). Its bimodal
form suggests to approximate a contact by a 2-state system, with rates
$\kclose$, $\kopen$ for binding and unbinding, respectively. To test whether
such a reduced description is sufficient, we initiate simulations in the fully
wrapped state and determine the functionally relevant time scales, i.e., the
average time $\tclose(n)$ until contact $n$ opens to expose the $n$th DNA
segment and the average time $\topen(n)$ until contact $n$ recloses
\cite{Allen:06,note3}. The results are shown in Fig.~\ref{fig2}(b) for $n\le 5$
\cite{note:estatic}. Within the reduced description of consecutive 2-state
contacts, $\tclose(n)$ can be calculated as a mean first passage time
\cite{Gardiner:83} for a 1D biased random walker with hopping rates $\kopen$,
$\kclose$. The walker starts at site zero (reflecting boundary) and reaches
site $n$ after an average time
\begin{equation}
  \label{MFPT}
  \tclose(n) = \frac{\kopen^{-1}}{1-\Keq}\left[\frac{1-\Keq^n}{1-\Keq^{-1}}+n\right] \stackrel{K\gg 1}
{\approx} \frac{\Keq^{n-1}}{\kopen}\;.
\end{equation}
Here, $\Keq=\kclose/\kopen$ can be interpreted as the effective equilibrium
binding constant per contact. The exponential increase of $\tclose(n)$ is clear
also from the equivalence of the biased random walk with a random walk against
a free energy ramp. The excellent fit of (\ref{MFPT}) to the simulation data
(dashed line) indicates that the reduced description is sufficient for the
dwell times in the inaccessible state. In contrast, it proves insufficient for
the dwell times in the accessible state, because $\topen(n)$ in
Fig.~\ref{fig2}(b) is clearly not constant as one would expect with a fixed
binding rate $\kclose$. Thus, we find $\topen(n)$ to be a more sensitive probe
for the physics of spontaneous site exposure than $\tclose(n)$. 

\begin{figure}[bt] 
\includegraphics[width=8.4cm]{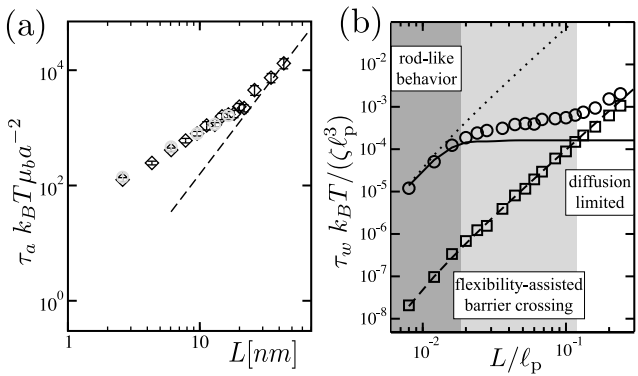}
\caption{\label{fig3} 
(a) The dependence of the dwell time $\topen(n=\!1\!)$ on the overhanging DNA length (diamonds) is compatible with $\topen(n)$ when $n$ is converted to contour length (gray circles). The dashed line indicates the diffusion limit (see main text for details). 
(b) The average barrier crossing time $\mwt$ (circles) for the SBR model of Fig.~\ref{fig1}(b). At small lengths, the barrier crossing time follows that of a stiff rod (indicated by the dotted line). Beyond a crossover length $\lc\ll\lp$, barrier crossing is much faster than for a stiff rod. For large lengths, $\mwt$ approaches the diffusion limit, i.e., $\mwt$ of the free SBR (squares). With $L<\lp$, free diffusion of the SBR (squares) is virtually indistinguishable from free diffusion of a rigid rod (dashed line). The crossover from the rodlike regime to the intermediate regime is well described by the theoretical analysis (solid line), see main text.}
\end{figure}

To probe the effect of the DNA length on the rewrapping kinetics, we vary the
number of overhanging beads before contact 1 and plot $\topen(1)$ as a function
of the overhang length $L$ in Fig.~\ref{fig3}(a). Superimposed is the data of
Fig.~\ref{fig2}(b) (bottom) with $n$ converted to contour length. The good
agreement of these dependencies indicates that $\topen$ is determined by
polymer dynamics. Indeed, we will now see that contact breaking and reformation
of a rotating semiflexible polymer displays much richer physics than a simple
1D barrier crossing process.

{\it Semiflexible Brownian Rotor.---}
The essential physics of contact formation in the nucleosome is captured by the
toy model depicted in Fig.~\ref{fig1}(b): A semiflexible polymer with contour
length $L$ and persistence length $\lp$ is attached to a point about which it
can rotate in a plane. The attachment angle $\varphi$ experiences a periodic
potential $V(\varphi)=V_{0}\cos(2\pi\,\varphi/\Delta\varphi)$, which creates
preferred angles separated by potential barriers as in our nucleosome model
(there, the barrier for contact reformation results from the DNA bending energy
and the electrostatic repulsion). The main difference is that the length of the
rotating polymer is constant for this `Semiflexible Brownian Rotor' (SBR),
while it changes slightly when a contact breaks or reforms in the nucleosome.
Also, we do not consider a directional bias in the SBR, because it is not
essential for what follows. So far, barrier crossing of semiflexible polymers
was studied only for situations where the entire polymer experiences an
external potential \cite{Kraikivski_EPL_04}. In the NCP, the potential acts
only on the angle at the attachment point. 

To characterize the phenomenology of the SBR, we determine its barrier crossing
rate $1/\mwt$ with Brownian dynamics simulations of a discrete bead-spring
model \cite{note:mwt}. The circles in Fig.~\ref{fig3}(b) show $\mwt$ as a
function of $L/\lp$ for $V_{0}=5\,k_{B}T$. We observe that at very short
lengths, $\mwt$ follows the stiff rod behavior $\mwt\sim L^{3}$ \cite{Doi:86}
indicated by the dotted line. However, above a certain length $\lc$, there is a
regime where $\mwt$ is nearly insensitive to $L$, before it rises again. Hence,
for lengths $L>\lc$ the semiflexible polymer crosses the barrier much faster
than the stiff rod. What is the physical mechanism for this acceleration? One
effect of a finite flexibility is a reduced mean end-to-end distance (due to
the undulations in the contour), which in turn leads to a larger rotational
mobility. However, with $V(\varphi)=0$, the rotational diffusion time of a
semiflexible polymer over an angle $\Delta\varphi$ (squares) is almost
identical to that of a stiff rod (dashed line) when $L<\lp$. Hence the
acceleration is not a mobility effect. Note that the dashed line is also the
diffusion limit for $\mwt$, which induces a second crossover from a reaction to
a diffusion controlled process. The equivalent diffusion limit is shown also in
Fig.~\ref{fig3}(a) (dashed line). It indicates that the $\topen(n)$ data for
the nucleosome is indeed in the accelerated barrier crossing regime.

{\it Flexibility-assisted barrier crossing.---}
To understand the interplay between the polymer dynamics and the barrier
crossing dynamics qualitatively, we recall the basic aspects of each: 
(i) A semiflexible polymer of length $L$ relaxes its conformational degrees of freedom in a time $\sim 
L^4/\lp$ \cite{LeGoff:02}. Conversely, within a given time $\tau$, a local bending deformation is 
``felt'' only over a length $\ell\sim(\lp\tau)^{1/4}$. 
(ii) The probability current over a barrier is proportional to the quasiequilibrium occupancy of the 
transition state and to the relaxation rate $\tau^{-1}$ out of this state. 
Together, (i) and (ii) imply that $\lc$ is the length of the polymer segment
that gets deformed during the relaxation process away from the potential peak.
We estimate $\lc$ by noting that the attachment angle relaxes according to
$\dot{\varphi}=-\mu(\lc)\,\partial V/\partial\varphi$, where $\mu(\lc)\sim
\lc^{-3}$ is the rotational mobility of the deformed segment. Hence,
$\tau\sim\lc^3(\Delta\varphi/2 \pi)^2/V_{0}$ and with
$\ell_c\sim(\lp\tau)^{1/4}$, we find 
\begin{equation}
  \label{Lc}
  \lc=C\,\lp\,\frac{k_BT}{V_0}\,\Big(\frac{\Delta\varphi}{2\pi}\Big)^2\;,
\end{equation}
where $C$ is a constant to be determined below. For lengths below $\lc$, the entire polymer is 
involved in the relaxation process, i.e., it behaves like a stiff rod.

\begin{figure}[bt] 
\includegraphics[width=8.5cm]{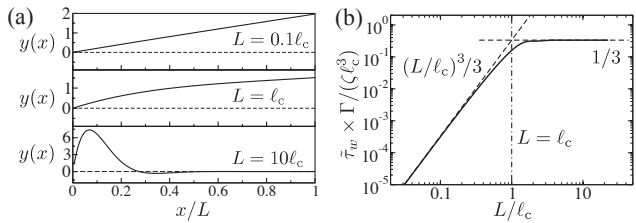}
\caption{Dynamics at the barrier. (a) The unstable eigenmode for three different lengths. Polymers shorter than $\Lcross$ rotate without significant deformation, while long polymers form a bulge of size $\sim\Lcross$ at the origin. (b) The prefactor of the Kramers time $\tilde{\tau}_w=1/\uev$ as a function of the length. The prefactor increases as $L^3$ if $L\ll \Lcross$ and is constant if $L\gg
\Lcross$.}
\label{fig:scaling}
\end{figure}

{\it Quantitative theory for the crossover.---}
To render the above picture quantitative, we employ the Langer theory for
multidimensional barrier crossing processes \cite{Langer:68}. For the case at
hand, one can show \cite{uslater} that the barrier crossing time simplifies to
$\mwt=\frac{\pi}{\uev}e^{2V_0/k_{B}T}$, where $\uev$ is the eigenvalue
associated with the unstable mode at the saddle point. We calculate $\uev$
using the continuous wormlike chain model in the weakly bending approximation
\cite{Wiggins:98}. At the transition state the chain is straight, e.g., along
the $x$ axis. We denote deviations from this configuration by $y(x,t)$. The
chain dynamics follows $\partial_t y=-(k_BT\lp/\zeta)\,\partial_x^4y$ with a
friction coefficient $\zeta$. With $\kruem=V_0(2\pi/\Delta\varphi)^2$ denoting
the curvature of the potential at the transition state, the torque on the
attached polymer end is $-\kruem\,\partial_x y|_{x=0}$. This torque must be
balanced by a local bend resulting in the boundary condition $k_BT\lp
\partial_x^2y|_{x=0}=-\kruem\,\partial_x y|_{x=0}$. The other boundary
conditions are $y|_{x=0}= \partial_x^2y|_{x=L}=\partial_x^3y|_{x=L}=0$. We find
a unique unstable mode with eigenvalue $\uev=k_ {B}T\lp\alpha^4/4\zeta L^4$ and
$\alpha$ determined by 
\begin{equation}
  \frac{\alpha[\sinh(\alpha)-\sin(\alpha)]}{\cosh(\alpha)+\cos(\alpha)+2} = \sqrt[3]{12}\,\frac{L}
{\lc} \;,
  \label{eq-for-k}
\end{equation} 
where $\lc$ is as in (\ref{Lc}) with $C=\sqrt[3]{12}$. In the limit $L\ll \Lcross$, we find $\uev=3
\kruem/\zeta L^3$ independent of the stiffness, whereas in the opposite limit $\uev=3\kruem/\zeta
\Lcross^3$ independent of $L$. \FIG{scaling} shows (a) the unstable eigenmode for $L/\lc=\{0.1,1,10\}$ 
and (b) the crossover in the barrier crossing time. The eigenmode shape confirms our qualitative 
picture: stiff and short polymers respond to the torque by rotating as a whole, whereas the torque 
shapes a bulge of size $\sim\lc$ in longer polymers. For a discrete polymer model, the same analysis 
can be performed, but the eigenvalue $\uev$ must be computed numerically. The solid line in Fig.~\ref
{fig3} shows the resulting barrier crossing time for the same discretization as used in the Brownian 
dynamics simulations of the SBR model. Indeed, the crossover from the rodlike
to the flexibility-assisted barrier crossing is well described by this
analysis. The deviations at larger $L$ can be attributed to finite barrier
corrections \cite{Talkner:94}.

{\it Discussion and outlook.---}
The experiments \cite{Li:05,Tomschik:05} have shown that the functionally
relevant time scales $\tclose$ and $\topen$ depend on the position on the
nucleosomal DNA. Our theoretical study suggests that these time scales
additionally depend on the total DNA length. The position dependence of
$\tclose$ should follow the random walker model (\ref{MFPT}), which is the
minimal model for a gradual, multistep opening mechanism. However, we expect
that the position dependence of $\topen$ and the length dependence of both time
scales will reflect the polymer dynamics of the DNA. Within our toy model, the
Semiflexible Brownian Rotor, we find three physically distinct regimes for this
length dependence, see Fig.~\ref{fig3}(b). The intermediate regime displays a
striking flexibility-assisted barrier crossing effect, the onset of which is
marked by the new length scale $\lc$ of Eq.~(\ref{Lc}). It can be interpreted
as the length over which the polymer contour is deformed as it passes over the
potential barrier. Because $\lc$ is considerably smaller than the persistence
length $\lp$, we expect that the onset of the intermediate regime will not be
detectable in nucleosomes. However, nucleosomes should display the crossover
from flexibility-assisted barrier crossing to diffusion-limited dynamics as
shown in Fig.~\ref{fig3}(a). All three regimes of Fig.~\ref{fig3}(b) could be
probed in an experimental realization of the SBR model, e.g., with an actin
filament as the rotating polymer. 

We thank H.~Boroudjerdi, T.~Franosch, E.~Frey, O.~Hallatschek, S.~Leuba,
R.~Netz, R.~Phillips, P.~Reimann, P.R.~ten~Wolde, and J.~Widom for useful
discussions, and the {\it DFG} for financial support.


\begin{thebibliography}{99}
\bibitem{note4}
Nucleosomal DNA is slightly unwound \cite{Luger:97}, with a small torsional energy $\sim 1\,k_BT$ per 10~bp. We include this twist contribution into the effective contact potential.
\bibitem{note2}
Note that our free energy cost per contact is slightly larger for the first contact and considerably 
larger when less than one turn of the superhelix remains wrapped, due to the electrostatic self-
repulsion of DNA.
\bibitem{note3}
We exclude transient wrapping or unwrapping events, i.e., we only count transitions that reach the 
metastable state in the neighboring potential valley.
\bibitem{note:estatic}
For $n>5$ the electrostatic self-repulsion of the DNA is less significant. This stabilizes the 
remaining contacts and modifies the exposure kinetics of the central sites (close to the dyad) \cite
{uslater}.
\bibitem{note:mwt}
The time $\mwt$ is operationally defined through the long-time rotational diffusion coefficient $D=\Delta\varphi^2/2\mwt$ over many potential wells. Thus, $\mwt$ measures the time for barrier crossing 
events that involve not only the passage of $\varphi$ to a neighboring valley in $V(\varphi)$, but also turning of the polymer tip into the direction of the new valley, analogous to $\topen$ in the nucleosome model \cite{note3}.
\end{thebibliography}
\end{document}